\documentclass[a4paper,12pt]{article}
\usepackage{setspace} 
\usepackage{footmisc}
    \newlength{\myfootnotesep}
\usepackage{amsmath}
\usepackage{amsthm}
\usepackage{amsfonts}
\usepackage{enumitem}
\usepackage{amssymb}
\usepackage{makeidx}
\usepackage{graphicx}
\usepackage[unicode=true,breaklinks=true,hidelinks]{hyperref}
\usepackage{csquotes}
\usepackage{setspace}
\usepackage{color}
\usepackage{subcaption}
\usepackage{tikz}
\usepackage{pgfplots}
\usepackage{float}
\usepackage{adjustbox}
\usepackage{hyperref}
\usepackage{epigraph}

\newcommand{\defeq}{\mathrel{\mathop:}=}

\usepackage[margin=0.8in]{geometry}
\graphicspath{ {images/} }
\usepackage{tikz}
\setlist[enumerate]{label*=\arabic*.}
\makeindex
\date{\vspace{-5ex}}
\begin{document}
\title{Spacetime Emergence and the Fear of Intimacy \\
\small \textit{penultimate draft}}
\author{Samuel T. Baron\thanks{University of Melbourne; \texttt{samuel.thomas.baron@gmail.com}}, \\
Claudio Calosi\thanks{University of Venice; \texttt{claudio.calosi@unive.it}}, \\
\& Cristian Mariani\thanks{Università della Svizzera Italiana, Lugano; \texttt{cristian.mariani@usi.ch}} }

\maketitle

\bigskip

\begin{abstract}
\noindent We provide a reply to the Argument from Intimacy on behalf of defenders of emergent spacetime in theories of quantum gravity. We argue that if one accepts that spacetime regions are nowhere in the sense that they are locations but do not have locations, then the Argument from Intimacy can be resolved. We go on to consider a problem with this response, namely that it is unavailable to super-substantivalists. We argue that this is right for {\em identity} but not {\em priority} super-substantivalists. We then suggest that there is no cost for our solution here, since identity versions of super-substantivalism face severe challenges in the context of spacetime emergence and so should be rejected anyway.
\end{abstract}

\section{Introduction}

According to several approaches to quantum gravity, spacetime is emergent. It is thus not a fundamental feature of reality.\footnote{By `spacetime' we mean a manifold equipped with the metric structure derived from the Einstein field equations.}$^,$\footnote{This claim has become particularly important in the context of the apparent threat of empirical incoherence (see e.g. Huggett \& Wüthrich 2013). This is a threat to all of theories of quantum gravity that do not feature any spacetime structure at all, even at the emergent, derivative level.} What is fundamental, if anything, is non-spatiotemporal. How we should understand the relevant notion of emergence when applied to the spacetime case is, of course, open to debate.\footnote{In the philosophy of physics literature, the goal appears to be to show that a given theory which describes relativistic spacetime can be appropriately \textit{derived} from a theory that does not feature any spacetime structure (for an overview, see Crowther 2018). Such a merely \textit{inter-theoretic} use of emergence, however, does little justice to the idea that what is described by the spacetime-free theory is \textit{more fundamental}, in a metaphysical sense, than what is described by the theory featuring spacetime. Although we are going to simply assume this without further argument, we note that it is intuitive to expect the inter-theoretic relation of emergence to mirror some metaphysical relation of priority in the world. For more metaphysical approaches to spacetime emergence see, e.g., Le Bihan and Baron (2021), Le Bihan (2021) and Lam \& Wüthrich (2018).} An influential, widely discussed argument---the Argument from Intimacy---allegedly instructs us as to how it should \textit{not} be understood.\footnote{See  e.g., Baron (2020, 2021) and Baron Miller \& Tallant (2022).} 

The Argument from Intimacy is predicated upon the assumption that composition is (at least) a necessary condition for emergence.\footnote{Some suggest it is also sufficient (Humphreys 1997). We cannot do justice here to the huge literature on emergence in metaphysics. We note, however, that on many ways of understanding the \textit{dependence} side of emergence (the other side being \textit{autonomy}), composition does indeed play an important role. Moreover, we note that the Argument from Intimacy can be formulated without assuming that composition is necessary for emergence. In particular, the argument can be formulated as an argument against the use of mereology to understand the connection between spacetime and a more fundamental, non-spatiotemporal reality, regardless of whether that connection is one of emergence and thus regardless of whether emergence requires mereology. For a comprehensive discussion of metaphysical emergence, see Wilson (2021).} The argument exploits an intimate relation between location and parthood, namely that wholes should inherit their locations from their parts, in the sense that they should be located wherever their parts are located. The basic idea behind the argument is thus to show that the inheritance of location fails when spacetime or spatiotemporal entities are taken to emerge from something non-spatiotemporal. Faced with the threat of such an argument---so the thought goes---one should abandon the mereological understanding of spacetime emergence and look somewhere else. 

In this paper, we offer a solution to the Argument from Intimacy. The solution involves rejecting the claim that spacetime regions are somewhere. Spacetime regions {\em are} spatiotemporal locations but do not {\em have} spatiotemporal locations. We explain why this solution is better than existing solutions to the Argument from Intimacy, before responding to three arguments in favour of the view that spacetime regions are somewhere. Two of these have been provided by Baron (2020), and are largely dealt with by the mereological framework for spacetime that we set out. The third focuses on super-substantivalism and forces us to say a bit more about substantivalist pictures in the context of spacetime emergence. 

We take our discussion of the Argument from Intimacy to be important for three reasons. First, what we say provides indirect support for existing work by Baron and Le Bihan (2021) and, particularly, Le Bihan (2018a, b) who defend mereological models of spacetime emergence. In particular, we show that a perfectly standard mereology can be used for mereological approaches to spacetime emergence. We thus make a case against the prevailing view that mereology must be `reimagined' to fit with the emergence of spacetime. This view is stated clearly by Baron, Miller and Tallant (2022, p. 136) when they write that:

\begin{quote}
... composition and constitution as standardly understood appear to be conceptually wedded to spatiotemporal notions. The standard way of understanding these two relations therefore makes them ill-suited to underwrite the dependence of spacetime on a non-spatiotemporal structure.
\end{quote}

Second, our discussion sheds light on the relationship between spacetime emergence on the one hand, and location and super-substantivalism on the other. In particular, we show that identity versions of super-substantivalism are at odds with spacetime emergence, whereas priority versions do better. We also show that  mereological approaches to spacetime emergence fit quite well with priority versions of super-substantivalism. 

Third, our arguments bear on a debate within the literature on formal theories of location about the location of spacetime regions. As we will see, the two main alternatives are that spacetime regions are located at themselves, or that spacetime regions are located nowhere. According to the orthodoxy, the choice between these two options is a matter of simple stipulation and, as such, is of no metaphysically substantive consequence, as e.g., both Varzi (2007: 1016) and Parsons (2007: 224) explicitly claim. The arguments in this paper challenge this orthodoxy, for we show that there are substantive metaphysical consequences related to the locations of spacetime regions after all.  

\section{Preliminaries}

Much of what we say in what follows makes use of mereological notions, and so it makes sense to render these precisely from the very beginning. This formal work will pay off later, since it will provide a straightforward framework within which to state the Argument from Intimacy more precisely than has been done so far, and to also outline our preferred solution. 

First: mereology. We assume a parthood relation $P$ that is transitive, reflexive, and asymmetric. We add the following further standard definitions:\footnote{These are indeed standard definitions, so we will not spend much time on them. See Cotnoir and Varzi (2021).}

\begin{description}
    \item \textsc{Overlap}: $O(x,y)$ $\defeq$ $\exists z (P (z,x) \wedge (z,y))$ 
    \item \textsc{Fusion}: $F_{\phi}z \defeq \forall x (\phi(x) \rightarrow P(x, z)) \wedge \forall y (P(x,z) \rightarrow \exists x (\phi(x) \wedge O(x,y)))$
\end{description}

Next: location. Although there are many distinct notions of location, we only need to look at those that are arguably the ones that have been central in almost any formal theory of location,\footnote{For a recent argument for the view that we should go beyond such notions see Correia (2022).} which include  \textit{Exact}, \textit{Weak}, \textit{Entire} and \textit{Pervasive} Location. For present purposes, we focus on exact and weak location.\footnote{
Intuitively, $x$ is entirely located at a region iff it lies within that region, and $x$ is pervasively located at  a region iff it completely fills that region. As an illustration, you are entirely located in the Milky Way, and you are pervasively located where your heart is.} To have a preliminary grasp on such notions, think of the ``exact location'' of an object $o$ as $o$'s shadow in substantivalist spacetime, as Parsons (2007) puts it. As for ``weak location'', this is the locative relation that an object $o$ bears to every spacetime region which is not completely free of $o$---to follow Parsons (2007) again.\footnote{See also Gilmore (2018).}   

It is usual to take one locative notion as primitive and then use it define the others.\footnote{For a theory that uses Exact Location as a primitive see Parsons (2007). For one that uses weak location see Eagle (2016, 2019). For one that uses entire location see Correia (forthcoming). Finally, for one that uses Pervasive Location see Loss (2021). For an argument that one may need more primitives see Kleinshmidt (2016).} We are going to use exact location, which we express with the two-place predicate $L(x,y)$.\footnote{But the argument works, \textit{mutatis mutandis}, for other choices of primitive as well. All formulas are intended to be universally closed.} Then we can define the other notions using $L$ and mereological predicates of parthood ($P$) and overlap ($O$).\footnote{See Casati and Varzi (1999), Parsons (2007: 204), Gilmore (2018).} We thus define weak location as follows:\footnote{We can define Entire and Pervasive location as follows:\begin{description}
    
    \item \textsc{Entire Location}: $EL(x,y)\defeq \exists z (L(x,z) \wedge P(y,z))$  
     \item \textsc{Pervasive Location}: $PL(x,y)\defeq \exists z (L(x,z) \wedge P(z,y))$  
\end{description}}

\begin{description}
    \item \textsc{Weak Location}: $WL(x,y) \defeq \exists z (L(x,z) \wedge O(z,y))$
\end{description}

\noindent We take the second argument of every locative notion to be a spacetime region. Formally, this can be done by, e.g., introducing a primitive predicate of ``being a spacetime region'' ($R$), and then adding the following axiom: 

\begin{description}
\item \textsc{Second Argument Spacetime Region}: $@L(x,y) \rightarrow R(y)$
\end{description}

\noindent Where $@L$ stands for any of weak, entire or pervasive location.

We can also define a spatiotemporal entity as an entity that is weakly located somewhere:

\begin{description}
\item \textsc{Spatiotemporal Entity}: $SP(x) \defeq \exists y WL(x,y) $
\end{description}

\noindent We will take spacetime regions to be a model of classical extensional mereology. This guarantees that any non-empty collection of regions has a unique fusion. Defining: 

\begin{description}
\item \textsc{General Fusion Operator}: $\sigma x (\phi(x)) \defeq \iota z \forall x (\phi(x) \rightarrow P(x,z)) \wedge \forall y (P(y,z) \rightarrow \exists x (\phi(x) \wedge O (x,y)))$
\end{description}
 
\noindent We can then take spacetime to be the fusion of all regions: 

\begin{description}
\item \textsc{Spacetime}: $\mathcal{S} \defeq \sigma z (\exists x (R(x) \wedge z = x)) $
\end{description}

As Varzi points out, there are two ways that spacetime regions can connect to location.\footnote{Here's Varzi: \begin{quote}
    There are two options here, depending on whether
we think that spatial regions are themselves entities located somewhere.
If we think so, then the obvious thing to say is that such entities can
only be located at themselves (Varzi, 2007: 1015).
\end{quote}

See also Casati and Varzi (1999: 121).} First, spacetime regions may, themselves, be spatiotemporally located. In the present context, this translates into the following:\footnote{As Varzi (2007) points out, \textit{if} regions have (exact) locations, the only plausible canndidate for an exact location of a give region is itself.}

\begin{description}
\item \textsc{Location of Regions}: $R(x) \rightarrow L(x,x)$ 
\end{description}

\noindent Note that it follows by definition of \textsc{Weak Location}, \textsc{Entire Location} and \textsc{Spacetime}, that regions are both weakly and entirely located in spacetime, if they are located at all. For now, we will just focus on the weakest of these locations for spacetime regions, namely:

\begin{description}
\item \textsc{Somewhere}: $R(x) \rightarrow \exists y (WL(x,y))$
\end{description}

\noindent Given \textsc{Spatiotemporal Entity} it follows that:

\begin{description}
\item \textsc{Regions are Spatiotemporal}: $R(x) \rightarrow SP(x)$
\end{description}

\noindent In effect, in the presence of \textsc{Location of Regions}, the two principles \textsc{Somewhere} and \textsc{Regions are Spatiotemporal} are equivalent, i.e., \textsc{Somewhere} $\leftrightarrow$ \textsc{Regions are Spatiotemporal} holds. At the other end of the spectrum, so to speak, there is the thesis that:

\begin{quote}
    [R]egions do \textit{not have a location}---they \textit{are} locations (Varzi, 2007: 1016, italics in the original).
\end{quote}

\noindent Not only this is a respectable position in the literature on formal theories of location. Simons (2004: 345) comes close to accepting the following as a non-negotiable axiom:

\begin{description}
\item \textsc{No Location for Regions}: $R(x) \rightarrow \neg \exists y (L(x,y))$ 
\end{description}

\noindent It follows that regions are neither weakly, nor entirely located anywhere. In particular, they are neither weakly nor entirely located in spacetime. Indeed we can set:

\begin{description}
\item \textsc{Nowhere}: $R(x) \rightarrow \neg \exists y (WL(x,y))$
\item \textsc{Regions are not Spatiotemporal}: $R(x) \rightarrow \neg SP (x)$
\end{description}

\noindent As before, in the presence of \textsc{No Location for regions}, the two principles \textsc{Nowhere} and \textsc{Regions are not Spatiotemporal} are equivalent, that is, \textsc{Nowhere} $\leftrightarrow$ \textsc{Regions are not Spatiotemporal} holds.

There are two things to note about \textsc{Nowhere} and \textsc{Somewhere}. First, as we've just seen, a perfectly precise mereological framework for spacetime location can be formulated using {\em either} principle. It is not the case, for instance, that only one of these notions can be rendered precise, and the other is somehow mysterious. Second,  \textsc{Nowhere} and \textsc{Somewhere} present a choice point in spatiotemporal mereology. One can hold the view that spacetime regions are, in addition to being parts of spacetime, also located in spacetime, thereby forcing spacetime itself to be located. Or, one can hold that spacetime regions are parts of spacetime, but not located in spacetime. As will become important later on, this second view---which involves endorsing \textsc{Nowhere}---is compatible with spacetime regions being spatiotemporal locations, it is just that they get to be this way by virtue of standing in a mereological relation to spacetime, rather than a locative one (as is required given \textsc{Somewhere}).

We need two more notions to set up the Argument from Intimacy. The first of these is the {\em inheritance of location}, which we take from work by Sider (2007). The basic idea behind the inheritance of location is that parts and wholes share their locations, in some sense. That is, what we don't generally see are cases in which the whole is located in a region and its parts are not also located within that region, or vice versa. According to Sider (2007), the inheritance of location forms part of our conceptual understanding of parthood. Part of what it is for a relation to be a parthood relation is that it respects the inheritance of location. As he puts it:

\begin{quote}
    Everyone accepts the inheritance principles. If they are true, then the part-whole connection is a uniquely intimate one. The intimacy of this connection must be explained. The best explanation is a conception of parthood that renders the connection between parts and wholes as intimate and identity-like as possible. (Sider 2007, p. 75)
\end{quote}

\noindent We will return to this point later on. For now, we will simply sharpen the principle as follows:\footnote{The principle stated here is what Baron (2021) calls the `downward' principle of inheritance, since it carries location `down' from wholes to parts. The `upward' principle carries location from parts to wholes, and can be stated as follows:
\begin{quote}
\textsc{Upward Inheritance of Location} $(P(x,y) \wedge L(x,z)) \rightarrow WL(y,z)$.  
\end{quote}
This appears to be the principle that Sider focuses on. Here's Sider's formulation:
\begin{quote}
\textbf{Inheritance of Location}: If $x$ is part of $y$, then $y$ is located wherever $x$ is located. (Sider, 2007: 25).
\end{quote}
As Baron (2021) shows, it is possible to formulate the Argument from Intimacy using the upward principle as well. However, the argument is much simpler when formulated using the downward principle, and so that is the principle we use here, but nothing of consequence hangs on this choice. 
}

\begin{description}
 \item \textsc{Inheritance of Location}: $(P(x,y) \wedge L(y,z)) \rightarrow WL(x,z)$.   
\end{description}

\noindent This says that if $x$ is part of $y$, then $x$ must be at least weakly located wherever $y$ is located, thus ruling out any cases in which a whole occupies a location that is disjoint from any of its parts. Next, we need a notion of {\em mereological spacetime emergence}. This is the idea, roughly, that spacetime is composed of non-spatiotemporal parts. This idea implies the following weak condition:

\begin{description}
 \item \textsc{Mereological Spacetime Emergence}: $\forall x (R(x) \rightarrow \exists y (Pyx \wedge \neg SP(y)))$
\end{description}

\noindent Strictly speaking, mereological spacetime emergence is a stronger view than the one stated, since it will likely require that each region is made of multiple parts. Such a condition could be stated in the language of plural logic, but there's no need to. The weak condition is sufficient to get the Argument from Intimacy going. 

\section{The Argument from Intimacy}

We are now in a position to state the argument from intimacy. We can state it as an inconsistent triad: 

\begin{description}

\item (1) \textsc{Mereological Spacetime Emergence}: $\forall x (R(x) \rightarrow \exists y (Pyx \wedge \neg SP(y)))$

\item (2) \textsc{Somewhere}: $R(x) \rightarrow \exists y (WL(x,y))$

\item (3) \textsc{Inheritance of Location}: $(P(x,y) \wedge L(y,z)) \rightarrow WL(x,z)$

\end{description}

\noindent Given \textsc{Mereological Spacetime Emergence}, every spacetime region $r$ has non-spatiotemporal parts, i.e. parts that are not spatiotemporally located. By the \textsc{Inheritance of Location}, the parts of a spatiotemporal region $r$ must be located wherever $r$ is located. By \textsc{Somewhere}, all spatiotemporal regions are spatiotemporally located. So the parts of a spatiotemporal region $r$ must all be spatiotemporally located. But, by \textsc{Mereological Spacetime Emergence}, some of those parts are not spatiotemporally located. So we have a contradiction. 

Assuming a commitment to {\sc Mereological Spacetime Emergence}, in order to resist the argument from intimacy one can either reject \textsc{Inheritance of Location} or reject \textsc{Regions are Somewhere} (or both). Existing responses to the argument recommend giving up \textsc{Inheritance of Location}.\footnote{Baron, Miller and Tallant (2022a) consider and reject this option. Baron and Le Bihan (2021) consider this option and partially endorse it, though they admit that it would be better if \textsc{Inheritance of Location} could be preserved.} We propose giving up \textsc{Somewhere} instead. 

However, matters are not quite so straightforward, for two main reasons. First, Baron (2020: 2214-2215) offers two arguments in favor of \textsc{Somewhere}. Both arguments must therefore be addressed. Second, and as we shall see, there is an argument in favour of \textsc{Somewhere} from super-substantivalism. Since super-substantivalism is a prominent account of the relationship between spacetime and matter, and one that has recently been discussed and defended exactly in the context of spacetime physics,\footnote{See among others Calosi and D\"{u}rr (2021), Gilmore (2014), and Lehmkhul (2018).} this argument must also be addressed.

Before moving on to consider the arguments for \textsc{Somewhere}, it is important to first address a prior question: why not just give up \textsc{Inheritance of Location}? Obviously, this would resolve the problem and, as noted, this is a response that some philosophers already favour. It is thus worth saying something here in favour of \textsc{Inheritance of Location}, although we note that our aim is not to provide a full-blown defense of the inheritance of location. All we need to show is that rejecting \textsc{Somewhere} is, all else being equal, a \textit{better} option than rejecting \textsc{Inheritance of Location}. And indeed we think this is the case for the following reason: while \textsc{Inheritance of Location} has been considered by many a non-negotiable constraint \textit{either} on parthood,\footnote{This claim is explicitly endorsed by Baron, Miller, and Tallant (2022), Cameron (2014), Gilmore (2010), Sider (2007), Leonard (2021b), Mellor (2008) and Schaffer (2009), among others.} \textit{or} on location, no similar claim has been made regarding \textsc{Somewhere}. Let us elaborate a bit on this.

There appear to be two distinct lines of reasoning that support \textsc{Inheritance of Location}. First, some believe that this principle is constitutive of the very notion of \textit{parthood} as a peculiarly {\em intimate} relation (see e.g., Cameron (2014) and Sider (2007)). The relation between a whole and its parts is clearly very different to the relation between a whole and other objects. One way to see this is to contrast parthood with other relations, such as causal relations, distance relations, and similarity relations. It seems clear that there is a deeper connection between a whole and its parts via the parthood relation than there is between the whole and other objects via (say) any causal relation. Indeed, in this respect parthood seems much more like identity than other relations: both identity and parthood forge a peculiarly strong connection between an entity and itself, a connection that is quite unlike the relations that an entity stands in to other entities. 

As Sider (2007, p. 54) discusses, the apparent similarity between identity and parthood underwrites a number of intuitive claims. These include: wholes are nothing over and above their parts; wholes just are their parts; wholes are no addition to being over and above their parts; wholes are an `ontological free lunch' paid for by their parts and so on. We find claims like this throughout philosophy. For instance, here's Lewis (1991, p. 81) on cats:

\begin{quote}
    But given a prior commitment to cats, say, a commitment to cat-fusions is not a {\em further} commitment. The fusion is nothing over and above the cats that compose it. It just {\em is} them. They just {\em are} it. Take them together or take them separately, the cats are the same portion of Reality either way.
\end{quote}

\noindent Similarly, Fine (2010, p. 572) writes:\footnote{
See also Varzi (1996, p. 264, fn. 2); Loss (2016, p. 489); and Smid (2017) for discussion. See Giordani and Calosi (2023) for a recent attempt to develop a mereology in which wholes are something above their parts.}

\begin{quote}
    ... a whole is a `mere sum'. It is nothing over and above its parts---or perhaps we should say, more cautiously, that it is nothing over and above its parts except insofar as it is one object rather than many.
\end{quote}

\noindent Indeed, the similarity between parthood and identity is one of the chief motivations behind the view that composition just \textit{is} identity.\footnote{Note that if composition just {\em is} identity, then \textsc{Inheritance of Location} follows via a very simply argument using Leibniz's Law. For suppose that the whole $w$ is identical to the parts $p_1 ... p_n$. Now, suppose that the whole is located at $l$ but that none of the parts are located at $l$. Then there seems to be a property that the whole has---namely, that of being located at $l$---that the parts lack. But then it follows that the whole is not identical to any of its parts, since they don't have the same properties. So, composition as identity is false.} Of course, as a number of philosophers have pointed out, composition as identity is a controversial view. But even if one rejects the view that composition is identity, this does not undermine the motivations that made this view seem plausible in the first place. That is, even if composition as identity is false, it still seems there is something plausible in the claim that the whole is nothing over and above the parts, that the whole just is the parts (though not the `is' of identity), that the whole is no addition to being and so on. But if there is any truth to these claims at all, then the inheritance of location looks quite plausible. For it is hard to see how the whole could be nothing over and above its parts, if (for instance) the whole has locative properties that are not carried by its parts. For then there would be quite a serious sense in which the whole is something over and above the parts.

This brings us to the second line of defence for \textsc{Inheritance of Location}. Even if \textsc{Inheritance of Location}is not constitutive of parthood, it is a crucial a principle for understanding location. The basic idea here is that we can do very little with the notion of location alone,\footnote{As a matter of fact, the only purely locative principles are \textit{Functionality} and \textit{Conditional Reflexivity}. See e.g. Casati and Varzi 1999: 121--123.} if we do not supplement it with mereological notions that allow us to connect it to parthood. The inheritance of location plays a crucial role in this regard, and it can therefore be considered non-negotiable as a principle in virtually any formal theory of location.\footnote{For instance, in Correia's (2022) theory---which is perhaps the most detailed, general theory of location produced to date---inheritance principles are partly axiomatic.} So much so, that one could even argue that we are allowed to reject certain philosophical views were they to conflict with \textsc{Inheritance of Location} (see Gilmore (2018)). 
Nothing remotely similar is true of \textsc{Somewhere}. The choice between \textsc{Somewhere} and \textsc{Nowhere} has no metaphysical significance for understanding location itself.

Here's a simple example to illustrate the point. Consider the definition of \textsc{Weak Location} given above. In this definition, \textsc{Weak Location} is tied to mereological overlap. Roughly, $x$'s weak location is defined to be any location that has a part $p$ in common with $x$'s exact location. Suppose, however, that \textsc{Inheritance of Location} is false. Then it is compatible with the fact that $x$'s weak location has a part $p$ in common with $x$'s exact location that $x$'s weak location corresponds to a location that is disjoint from the location of $p$ and thus from $x$'s exact location. But that means that part-sharing does not guarantee any substantive locative relationship between $x$'s weak location and $x$'s exact location. However, the whole point of defining $x$'s weak location in terms of part-sharing was to guarantee such a locative relationship. This is needed to capture the intuitive picture of what a weak location is, namely something that shares in an exact location, without being identical with it. Now, perhaps in response one could simply do away with weak location, or define it without mereological notions. However, the first option leaves the notion of location severely underpowered when it comes to applications, and the second option (to our knowledge) has never been done successfully.

In sum, then, compared to rejecting \textsc{Inheritance of Location}, rejecting \textsc{Somewhere} poses less of an epistemic risk. By rejecting \textsc{Inheritance of Location}, one runs the risk of giving up on our basic understanding of parthood,which in turn risks directly undermining the idea that spacetime emerges by virtue of having non-spatiotemporal parts.\footnote{Granted, one can still invoke a relation with certain formal properties (namely those that are specified by the axioms of classical mereology), but these formal properties would not count as parthood in any meaningful sense. Moreover, these formal properties on their own are not very explanatory. All they do, effectively, is specify a certain type of partial order. Since what we want is an explanation of how spacetime emerges from an underlying non-spatiotemporal metaphysics, we need to go beyond formal properties toward metaphysical features that shed light on the situation.} In addition, one runs the risk of undermining any viable theory location. Without a viable theory of location, however, it is not clear to what extent one can talk about even spatiotemporal locations any more except, perhaps, in a very weak sense. By contrast, the rejection of \textsc{Somewhere} does not carry these risks. Of course, if \textsc{Somewhere} is supported by powerful arguments, then that would alter the balance sheet. But, as we shall now argue, that doesn't seem to be the case. 

\section{Two Arguments for \textsc{Somewhere}}

We turn now to the discussion of Baron's two arguments in favour of \textsc{Somewhere}. Let us quote Baron at length:

\begin{quote}
   One might deny that spacetime regions are located within spacetime. [...] But the picture is unsatisfactory, for two reasons. First, it is difficult to understand how it is that spacetime regions fail to be located within spacetime. If regions are not located in spacetime, then they should not have any spatiotemporal properties. But they clearly do [...] Second, and relatedly, if spacetime regions are not located within spacetime, then any object that is located at a spacetime region would not, itself, be located in spacetime. For instance, Sydney is located at a certain spatiotemporal region $r$. Despite this, Sydney would not be located within spacetime because while it is located at $r$, that region itself is not located anywhere in spacetime. Denying this implication and maintaining that something is located in spacetime by being located in regions that are not located in spacetime is a bit like saying that Sara is located in France because she is in Nice despite the fact that Nice is not in France. Regions need to be located within spacetime in order to make sense of generic, object location within spacetime (2020: 2214-2215)
\end{quote}

\noindent According to Baron's first argument, \textsc{Somewhere} is needed to account for spatiotemporal properties of regions. If spatiotemporal regions are not located in spacetime they can have no spatiotemporal properties, but they clearly do. According to the second argument, \textsc{Somewhere} is needed to make sense of the location of material objects in spacetime. If regions are not located in spacetime, an object located at a given region cannor be located in spacetime either, but it clearly is. Let us call the first argument the {\em Spatiotemporal Properties Argument} and the second argument the {\em Location of Objects Argument}. 
In what follows we shall start by addressing the Location of Objects argument. This is because our response to it provides the basis to address the Spatiotemporal Properties argument as well. 

Within the framework laid out in \S2, the Location of Objects Argument amounts to the following claim: \textsc{Somewhere} is needed to account for the fact that material objects are located in spacetime. Given the distinction we drew between weak and entire location, this claim might be translated in at least two different ways, where $O$ stands for ``being a material object'':\footnote{Clearly, Exact and Pervasive Location would not be satisfactory choices.}

\begin{description}
\item \textsc{Objects in Spacetime}$_{W}$: $O(x) \rightarrow WL(x, \mathcal{S})$
\item \textsc{Objects in Spacetime}$_{E}$: $O(x) \rightarrow EL(x, \mathcal{S})$
\end{description}

\noindent These principles simply say that material objects are weakly or entirely located in spacetime respectively. Thus, Baron's claim is that if \textsc{Somewhere} is false, then we must also reject both of the claims just stated. But that is simply not true. Consider \textsc{Objects in Spacetime}$_{W}$ and suppose that \textsc{Somewhere} is false. In order for \textsc{Objects in Spacetime}$_{W}$ to be true, either there are no material objects, or all material objects are weakly located in spactime. Clearly there are material objects, so it is the second disjunct that must be defended. Given the mereological framework set out in \S2, however, all it takes for an object to be weakly located in spacetime is for that object to be located at a spatiotemporal region. We do not also require that spatiotemporal regions are themselves located in spacetime. 

To see this a bit more clearly, it is important to consider the way in which spacetime was constructed in \S2. There we began with a primitive predicate, that of being a spatiotemporal region. Spacetime is then just defined as the mereological fusion of all such regions. As we also saw, it is compatible with this construction that spatiotemporal regions are themselves not {\em located} in spacetime, since one is free to endorse \textsc{Nowhere} without in any way undermining the idea that spatiotemporal regions are parts of spacetime, or that they provide locations for spatiotemporal entites. As a result, material objects can be located in spacetime by virtue of being located at parts of spacetime, even though those parts are not themselves spatiotemporal located. If that's right, though, then location of regions is not needed to account for the location of objects in spacetime. All that is needed is that regions stand in a \textit{mereological} relation to spacetime, not a \textit{locative} one.\footnote{We used mereology, as is customary in the literature on formal theories of location, but it is important to note that the argument generalizes. One can define Weak Location and Entire Location in terms of Exact Location and other relations. Indeed, let us use $E$ for a general relation of ``Being an element of''. Examples of $E$- relations include---but might not be limited to---\textit{parthood}, \textit{subsethood}, the relation of \textit{being one of} familiar from plural logic, and the like. Then we could define Weak and Entire Location using $E$---and notions defined in terms of $E$. For the sake of illustration, here is Entire Location$^*$:

\begin{description}
    
    \item \textsc{Entire Location}$^*$: $EL(x,y)\defeq \exists z (L(x,z) \wedge E(y,z))$  
\end{description}
 
One obtains different notions of Entire Location by replacing $E$ with different, specific $E$-relations. The argument we gave can be formulated in terms of the general relation $E$. This strengthens the conclusion: all one needs to account for the location of objects in spacetime is that spacetime regions and spacetime are $E$-related. Spacetime regions and spacetime need not be related by \textit{any} locative relation. Needless to say, to formulate a proper counterpart of the argument in the main text one needs to give all relevant definitions---e.g., that of spacetime---in terms of the chosen $E$-relation.}
 
As anticipated, this line of argument also undermines the Spatiotemporal Properties argument. To recall, \enquote{[I]f regions are not located in spacetime, then they should not have any spatiotemporal properties} (Baron, 2020: 2215). But the argument above suggests that spacetime regions may derive the relevant spatiotemporal properties by being parts of spacetime, rather than being located in it. So, for example, a region can have certain metric properties simply by virtue of being a part of a larger structure, which has those properties. This is in line with structuralist pictures of spacetime more generally. Spacetime is a structure, and regions are embedded parthood-wise within that structure and, in this way, enjoy the benefits of being within spacetime, such as the possession of spatiotemporal properties.

Note that, if need be, we can generalise this thought a bit. Let spatiotemporal regions not even be parts of spacetime. From a set-theoretic perspective, they may still be elements of it. But consider that it is unproblematic to account for metric properties (such as, e.g. \textit{being 3 meters apart}) in set-theoretic terms. Indeed, this is what we usually do when we define topological, metric, and even measurable spaces on a given set-theoretic structure. We never invoke locative notions. Nor do we need to. It appears safe to claim that locative notions were never considered essential for such an explanation.

\section{Super-substantivalism}

So far we have addressed two arguments in favour of \textsc{Somewhere}. We come now to the third. The argument, in brief, is just this: rejecting \textsc{Somewhere} conflicts with a family of views about the relationship between spacetime and matter, namely super-substantivalism. As we pointed out, this has been suggested as an ontological lesson to be drawn from our best spacetime physics---in particular from general relativity. Thus, it is important to address the argument. There are different variants of super-substantivalism that cash out the relation between spacetime and matter differently. According to one prominent view, this relation is simply identity (see, for instance, Leonard (2021a) and Schaffer (2009)). Using the formal machinery offered above, we can state a number of such views:

\begin{description}
\item \textsc{Identity Super-Substantivalism$_1$}: $O(x) \rightarrow \exists y (R(y) \wedge x = y))$
\item \textsc{Identity Super-Substantivalism$_2$}: $L(x,y) \leftrightarrow (O(x) \wedge R (y) \wedge x = y$)
\item \textsc{Identity Super-Substantivalism$_3$}: $SP(x) \rightarrow \exists y (R(y) \wedge x = y)$
\item \textsc{Identity Super-Substantivalism$_4$}: $L(x,y) \leftrightarrow (SP(x) \wedge R (y) \wedge x = y))$

\end{description}

\noindent Depending on whether one thinks that spatiotemporal entities are exhausted by objects, and whether objects have exact locations one gets that the views are interestingly related but not equivalent. We can  restrict our attention to \textsc{Identity Super-Substantivalism$_1$}: $O(x) \rightarrow \exists y (R(y) \wedge x = y))$ and
\textsc{Identity Super-Substantivalism$_2$}: $L(x,y) \leftrightarrow (O(x) \wedge R (y) \wedge x = y$), insofar as they are weaker---in general---than their $SP$-counterparts. 

Now, suppose one accepts \textsc{Identity Super-Substantivalism$_1$}. Suppose, in addition, that one accepts the principle that there is at least one object:

\begin{description}
\item \textsc{Non Emptiness$_O$}: $\exists x O(x)$
\end{description}

\noindent Then it follows that a weakened version of \textsc{Somewhere} is true. According to this weakened principle, there is at least one spacetime region that is spatiotemporally located. We can capture this principle by simply weakening \textsc{Somewhere} as follows:

\begin{description}
\item \textsc{Weak Somewhere}: $\exists x (R(x) \wedge \exists y (WL(x,y)))$
\end{description}

\noindent To see why \textsc{Weak Somewhere} follows from \textsc{Non Emptiness$_O$} and \textsc{Super-substantivalism$_1$} consider that if \textsc{Super-substantivalism$_1$} is true, then every object $x$ is identical to some spatiotemporal region $r$. Suppose then that there is some such entity $x$ as per \textsc{Non Emptiness$_O$}. Then there is some region $r$ such that $x = r$. As we said, we take objects to be spatiotemporal entities which are weakly located in spacetime by definition. Thus, because $x = r$, it follows that $r$ is weakly located as well, on pain of violating Leibniz's law. It thus follows that \textsc{Weak Somewhere} is true: there's at least one spatiotemporal region that is weakly located.

The trouble then is that \textsc{Weak Somewhere} can be used to reformulate the Argument from Intimacy. As before, the argument can be formulated as an inconsistent triad with \textsc{Inheritance of Location} and \textsc{Mereological Spacetime Emergence}: 

\begin{description}
\item (1) \textsc{Mereological Spacetime Emergence}: $\forall x (R(x) \rightarrow \exists y (Pyx \wedge \neg SP(y)))$

\item (2) \textsc{Weak Somewhere}: $\exists x (R(x) \wedge \exists y (WL(x,y)))$

\item (3) \textsc{Inheritance of Location}: $(P(x,y) \wedge L(y,z)) \rightarrow WL(x,z)$.
\end{description}

\noindent These three claims are inconsistent. By \textsc{Weak Somewhere} there is some region $r$ that is spatiotemporally located. By \textsc{Mereological Spacetime Emergence}, it follows that $r$ is composed of non-spatiotemporal parts $x_1, x_2, ..., x_n$, because all regions are. By \textsc{Inheritance of Location}, the $x_n$ are located wherever $r$ is located. But $r$ is located in spacetime. So the $x_n$ are located in spacetime. But they're not located in spacetime. So we have a contradiction.

It should also be noted that at least some versions of identity super-substantivalism \textit{cannot} reject \textsc{Inheritance of Location}. Thus, for these views, the second iteration of the Argument from Intimacy is even more serious than the first, since there really is no way out. This is best appreciated by considering a recent argument in Leonard (2021a). Leonard argues that those who endorse \textsc{Super-substantivalism$_{2(4)}$} are better off endorsing what he calls the Identity Theory of Location:\footnote{Leonard uses ``generalized identity'' in his formulation, but this will not matter for our purposes.}

\begin{description}
    \item \textsc{Identity Theory}: $\lambda x \lambda y (L(x,y)) := \lambda x \lambda y (O(x) \wedge R(y) \wedge ( x = y ))$
\end{description}

\noindent Roughly, this is just the view that exact location is identity. Simple proofs deliver that weak location is overlap, entire location is parthood, and pervasive location is the converse of parthood, extension.\footnote{These are left to the reader.} Abusing notation we have:

\begin{description}
    \item \textsc{WL-Overlap}: $WL(x,y) := O(x,y)$
    \item \textsc{EL-Parthood}: $EL(x,y):= P(x,y)$
    \item \textsc{PL-Extension}: $PL(x,y):= P(y,x)$
\end{description}

\noindent \textsc{Inheritance of Location} now boils down to the claim that parthood entails overlap---which is a trivial mereological principle. 

Perhaps, then, we should just reject super-substantivalism, deeming this to be the price of resolving the Argument from Intimacy without giving up \textsc{Inheritance of Location}. Unfortunately, this is not really an option for us. If one gives up super-substantivalism {\em tout court}, then one is left open to yet another iteration of the Argument from Intimacy. To see this, however, we need to take a step back and consider the relationship between matter and spacetime in the context of spacetime emergence.

As noted in \S1, the idea that spacetime is emergent comes from work in quantum gravity. In the context of quantum gravity, it is expected that both spacetime and spatiotemporally located entities---which we can think of as matter fields---will be emergent. The clearest indication of this comes from general relativity. In general relativity, the metric and the matter fields are coupled. It is plausible, then, that if spacetime is emergent, then so too are the matter fields within that theory. This suggests that the spatiotemporal description of matter is unlikely to be the most basic description. Because some theories of quantum gravity paint a picture of a fundamentally non-spatiotemporal world, it seems that the spatiotemporal description of matter must be emergent from something that is not spatiotemporal. Quantum mechanics is also expected to be emergent from a more basic theory of quantum gravity. The description of spatiotemporally located material entities such as quantum fields that we find in, for instance, quantum field theory is thus likely to be emergent. Again, if the more basic theory is not spatiotemporal, then this suggests that the the spatiotemporal description of matter should give way to a non-spatiotemporal picture of matter at some point. 

Now, if we accept super-substantivalism, then when we account for the emergence of spacetime, we thereby get the emergence of matter for free. That's because, on such a view, matter is sheeted home to regions of spacetime. If, however, super-substantivalism is rejected, then matter and spacetime come apart. Accordingly, a separate account of the emergence of matter must then be given, in addition to the account offered of the emergence of spacetime.\footnote{Of course, the situation is not the same if one were to endorse \textit{relationism}, for in that case too, arguably, the emergence of matter and spacetime would go hand in hand (by making matter emerge, we would get the emergence of spacetime for free).} In particular, what we need is an account of how material, spatiotemporal objects emerge from an underlying non-spatiotemporal metaphysics. If, however, we invoke a mereological approach to the emergence of matter as well as spacetime, then the Argument from Intimacy comes back in a nasty form. This version of the argument can be formulated as the following inconsistent triad (note that we have switched to the more general notion of spatiotemporal entity, because the notion of ``matter'' as in ``matter fields'' may comprise more than what we ordinarily think of as material \textit{objects}): 

\begin{description}

\item (1) \textsc{Entity Mereological Emergence}: $SP(x) \rightarrow \exists y (Pyx \wedge \neg SP(y))$.

\item (2) \textsc{Spatiotemporal Entities are Somewhere}: $SP(x) \rightarrow \exists y WL(x, y)$.

\item (3) \textsc{Inheritance of Location}: $(P(x,y) \wedge L(y,z)) \rightarrow WL(x,z)$.

\end{description}

\noindent \textsc{Entity Mereological Emergence} is just the claim that some spatiotemporal entities have parts that aren't spatiotemporally located. \textsc{Spatiotemporal Entities are Somewhere} is just the definition of a spatiotemporal entity given in \S2. These three claims are inconsistent. To see this, suppose that there is at least one spatiotemporally located entity, $e$. By \textsc{Entity Mereological Emergence}, $e$ will have parts $x_1, x_2, ..., x_n$ that are not spatiotemporally located. By \textsc{Inheritance of Location}, however, the $x_n$ will be located wherever $e$ is located. Since $e$ is spatiotemporally located, then so are the $x_n$. But at least one of the $x_n$ is not spatiotemporalyl located anywhere. So we have a contradiction: there is at least one part of $e$ that both is and is not spatiotemporally located. 

Notice that rejecting \textsc{Somewhere} won't help resolve this version of the argument. For even if spatiotemporal regions are not located, if spatiotemporal entities are located, then the problem remains. Note further that rejecting the claim that spatiotemporal entities are located won't help either. For it is built into the very definition of such entities that they have a spatiotemporal location. There are then just two options available for solving the third version of the Argument from Intimacy: either give up on the mereological approach to the emergence of matter, or give up on the Inheritance of Location. 

Neither option is a disaster, but they both seem unattractive. We have already explained why giving up on \textsc{Inheritance of Location} should be avoided if possible: because it is epistemically risky. The trouble with giving up on a mereological approach to the emergence of matter is that one must then provide something in its place. This makes the picture of emergence unpleasantly disjunctive: two radically different accounts are needed to explain the emergence of spacetime, on the one hand, and the emergence of spatiotemporally located material objects, on the other hand. A picture that uses just one approach to emergence for both would have the virtues of simplicity and elegance. 

Even if these considerations are not decisive, they do tend to weaken our case for rejecting \textsc{Somewhere} over \textsc{Inheritance of Location}. Our case, recall for why we should reject \textsc{Somewhere} is that doing so carries no real metaphysical risks, whereas the same cannot be said for giving up on \textsc{Inheritance of Location}. But that is much less plausible if it turns out that rejecting \textsc{Somewhere} prevents one from adopting certain pictures of the emergence of matter. Indeed, it would seem that the two options are then rather matched for the risks they take. We thus require an alternative solution to the problem at hand. What we need, in particular, is a way to render the rejection of \textsc{Somewhere} compatible with super-substantivalism. As we shall now argue, this is indeed possible. 

Recall, first, that the type of super-substantivalism used to formulate the problem is {\em identity} super-substantivalism. But this is not the only version of super-substantivalism available. An alternative version of super-substantivalism is the ontological priority view. On this view, spatiotemporal regions are ontologically prior to, but neither identical with nor parts of, spatiotemporal entities.\footnote{One alternative to the identity view is the composition view. On this view, material objects are composed of, rather than identical to, spacetime regions (see e.g., Gilmore (2014)). This version of super-substantivalism also requires that spacetime regions are somewhere in order to preserve \textsc{Inheritance of Location}. To see this, suppose that we have some material object $o$. According to the composition view, $o$ will be composed of spatiotemporal regions $r_1, r_2, ... r_n$. Suppose that spatiotemporal regions are nowhere, and thus have no spatiotemporal locations. Then it follows that $o$, which has a spatiotemporal location, is composed of parts, the $r_n$, which have no spatiotemporal location. This conflicts with the \textsc{Inheritance of Location}. For it turns out that $o$ is not located wherever its parts are located: $o$ is located in spacetime, but none of its parts are. Thus, the super-substantivalist is forced to accept that spacetime regions have spatiotemporal locations in order to preserve \textsc{Inheritance of Location}. They cannot then give this claim up in order to accommodate the idea that spatiotemporal regions are composed of non-spatiotemporal parts.} Exactly how to understand `ontological priority' is controversial, though it is usually thought to imply relative fundamentality.\footnote{For a discussion see Calosi and D\"{u}rr (2021).} Not to prejudge the issue let us consider a general determination relation $D$---which might include grounding, ontological dependence, and the like\footnote{It is an interesting and open issue whether all such determination relations will obey some kind of inheritance principle. While it appears safe to expect relations such as constitution to do it, it is much less obvious whether this is the case, for instance, for grounding---though see Baron, Miller and Tallant (2022) for a recent argument in this direction. Even if grounding {\em does} obey harmony, however, no-one has shown that this is true for all D-relations, and so we are confident that the view we call {\textsc{Super-duper Priority Substantivalism} does not face an obvious revenge argument from intimacy.}}---that tracks ontological priority. By that we simply mean that the following conditional holds ($>_f$, to be read as ``more fundamental than''):

\begin{description}
    \item \textsc{Tracking}: $xDy \rightarrow x >_f y$
\end{description}

\noindent An important feature of any such view is that it rejects any notion of identity between material objects and spacetime.\footnote{Indeed, $>_f$ is a (partial) strict order, if any.} That being so, there is no pressure toward {\textsc{Somewhere} of the kind already outlined for identity super-substantivalism. This means that a priority version of super-substantivalism is immune to the second iteration of the Argument from Intimacy. It also blocks the third: if matter ontologically depends on spacetime, then it is reasonable to suppose that matter comes ``for free''---thanks to the $D$-relation---with the emergence of spacetime. Moreover, because spacetime has metaphysical priority, it emerges first. What this means is that it is lower down on the chain of being compared to matter. As such, there is no need to mereologically compose material objects out of non-spatiotemporal entities. Rather, all of the composition can happen between spacetime regions and their non-spatiotemporal parts. Once composed, spatiotemporal regions can then provide the ontological basis for material objects, in the matter imagined by the priority view.

In recent work, Lehmkuhl (2018) argues in favour of the priority view over the identity view. His argument, roughly put, is that general relativity favours a priority interpretation of super-substantivalism compared to an identity interpretation. There is thus some reason to suppose that the identity version of super-substantivalism should be given up anyway. We note this as a point in our favour, but our aim is not to directly defend the priority interpretation of super-substantivalism here. Instead, we'll argue that the identity interpretation of super-substantivalism is a poor fit for the case of spacetime emergence; only the priority version will do. Thus, the only viable version of super-substantivalism available in the case of spacetime emergence is the very version that is compatible with rejecting \textsc{Somewhere}. 

To prepare the way, it is important to see that super-substantivalism must be reconfigured to align with spacetime emergence {\em anyway}. Super-substantivalism takes spacetime to be a substance. Taking spacetime to be a substance, however, implies that spacetime is fundamental. But this is precisely what we should reject if spacetime is emergent. Thus, super-substantivalism must be modified: a replacement substance for spacetime is needed. An account must then be given of the relationship between matter and whatever substance replaces spacetime.\footnote{Incidentally, this undermines a claim in Lehmkhul (2018) to the point that identity supersubstantivalism of this variety is immune from developments from physics.}

It makes sense to link the first point to physics. The new substance for super-substantivalism is whatever fundamental structure (sometimes called \textit{pre-matter} or {\em pre-geometry}) gets identified by a theory of quantum gravity. So, for instance, if loop quantum gravity turns out to be the correct theory of quantum gravity, then the fundamental substance is likely to be a spin-foam. If, by contrast, causal set theory is correct, then the fundamental substance is a causal set. Because we don't know much about this fundamental yet, let us just call it \textsc{substance$_{QG}$} and leave it open. 

In order to link matter to the \textsc{substance$_{QG}$}, we can just extend the two options for developing super-substantivalism already considered. This gives us two views to consider:

\begin{description}
    \item \textsc{Super-duper Identity Substantivalism}: Material objects are identical to \textsc{substance$_{QG}$}.
    \item \textsc{Super-duper Priority Substantivalism}: Material objects are less fundamental than \textsc{substance$_{QG}$}, and are \textit{D}-related to it.
\end{description}

\noindent \textsc{Super-duper Identity Substantivalism} faces an insurmountable problem. Assume, as seems plausible, that at least some material objects are located in spacetime. Such objects therefore have spatiotemporal properties. At the very least, they are weakly located at spatiotemporal regions. For instance, a tree is one such object, and it is clearly spatiotemporally located. Indeed, the whole point of taking spacetime to be emergent is to preserve intuitive claims of this kind---this is reflected in our characterization of a spatiotemporal entity (i.e. $SP(x)$).

However, note that \textsc{substance$_{QG}$} is supposed to be non-spatiotemporal and thus entirely lacking in any spatiotemporal properties. Any identification of spatiotemporally located entities with \textsc{substance$_{QG}$} would thus amount to a straightforward violation of Leibniz's law. For suppose that we identify a tree, which has spatiotemporal properties, with some part of the \textsc{substance$_{QG}$}, which has none. Then Leibniz's law is violated: the tree and the \textsc{substance$_{QG}$} have different properties, despite being numerically identical. Note that matters are not helped by {\em first} identifying material objects with regions of spacetime and then identifying spacetime with the \textsc{substance$_{QG}$}. For this would, once more, force spatiotemporal properties (e.g., being spacetime) down into the \textsc{substance$_{QG}$}, thereby violating Leibniz's law again.

\textsc{Super-duper Priority Substantivalism} avoids this problem entirely. Material objects are not identical to the \textsc{substance$_{QG}$} but only less fundamental, and related to it by some relation of determination (what we called earlier the \textit{D}-relation, which is tracking fundamentality). Without identity there is then no presumption that the properties possessed by material objects will be also possessed by the \textsc{substance$_{QG}$}. It is thus entirely open that material objects have spatiotemporal properties by virtue of being located in spacetime, whilst also being dependent on a \textsc{substance$_{QG}$} that has no such properties. Note also that making material objects directly depend on spacetime and then, indirectly, depend on the \textsc{substance$_{QG}$} makes no difference. For there is also no presumption that the properties of spacetime will be possessed by the \textsc{substance$_{QG}$} that it depends upon, and so there is no sense in which the dependence of spacetime on the \textsc{substance$_{QG}$} violates Leibniz's law.

We thus have a decisive case against \textsc{Super-duper Identity Substantivalism}. This shows that forms of super-substantivalism that require \textsc{Somewhere} should be rejected \textit{anyway}, once such views are formulated in a context where spacetime is an emergent feature. We should, instead, adopt a form of super-substantivalism that is entirely compatible with the rejection of \textsc{Somewhere}, and thus compatible with our solution to the Argument from Intimacy. This overcomes the problems stated above, and shows that there is no case to be made for \textsc{Somewhere} based on a commitment to super-substantivalism.

\section{Conclusion}

It is time to take stock. The Argument from Intimacy relies on the claim that spacetime regions are located in spacetime. As we have shown, however, this view is not forced upon us, for we could endorse \textsc{Nowhere}, and take spacetime regions to lack spatiotemporal locations. Following this, we identified a problem for this solution based on super-substantivalism. We resolved this problem by showing how to resolve the Argument from Intimacy in a manner that is compatible with the truth of super-substantivalism. 

We close by drawing out three points from the preceding discussion. First, there is no need to adopt a non-standard mereology to account for spacetime emergence. Instead, a standard mereological framework can be used, one that upholds \textsc{Inheritance of Location}. This shows that there is nothing particularly strange about a mereological approach to spacetime emergence. For all intents and purposes, it is just like standard cases of emergence, insofar as they involve relations between parts and wholes. This promises to have substantial explanatory pay-offs, since it means that we can extend our understanding of how emergence works in other cases, to the emergence of spacetime.

Second, our discussion has revealed a natural pairing between mereological approaches to spacetime emergence and \textsc{Super-duper Priority Substantivalism}. This is for three reasons. First, giving up super-substantivalism while retaining a mereological approach to emergence, forces one back into a version of the Argument from Intimacy and so the proponent of a mereological approach to spacetime emergence has some admittedly defeasible reason to be a substantivalist. Second, identity versions of super-substantivalism do not generalise well to the case of spacetime emergence, priority views are better, and so insofar as one is going to be a substantivalist, one should adopt a priority view. Finally, \textsc{Super-duper Priority Substantivalism} is compatible with rejecting \textsc{Somewhere} and, as we've noted, doing so resolves the Argument from Intimacy. Mereological approaches to spacetime emergence and \textsc{Super-duper Priority Substantivalism} are thus natural bedfellows.\footnote{Note we are not making the stronger claim that one view entails the other; one may still find space to be a relationist or a super-substantivalist of another kind.}

Our third general point relates to mereology. As noted, according to Varzi, the choice between the two principles below can be made simply via stipulation, without metaphysical consequence:

\begin{description}
\item \textsc{Somewhere}: $R(x) \rightarrow \exists y (WL(x,y))$

\item \textsc{Nowhere}: $R(x) \rightarrow \neg \exists y (WL(x,y))$
\end{description}

\noindent We have shown this to be false: endorsing \textsc{Nowhere} provides at least a first-pass solution to the Argument from Intimacy. The choice is thus not innocent at all: it has substantial implications for how we theorise about the emergence of spacetime. 


\section*{References}
\begin{description}
    \item Baron, S., 2020. “The Curious Case of Spacetime Emergence", {\em Philosophical Studies}, 177: 2207–2226.
    \item Baron, S. , 2021. ``Parts of Spacetime'', \textit{American Philosophical Quarterly}, 58 (4): 387-398.
    \item Baron, S., Miller, K. and Tallant, J., 2022a. {\em Out of Time; A Philosophical Study of Timelessness}. Oxford: Oxford University Press.
   \item Baron, S., Miller, K. and Tallant, J., 2022b. ``The Harmony of Grounding'', {\em Philosophical Studies}, 179: 3421-3446.
    \item Baron, S. and Le Bihan, B., 2022, ``Composing Spacetime'', {\em Journal of Philosophy}, 119(1): 33--54.  
    \item Calosi, C. Forthcoming. ``The One Magic Wave: Quantum Monism Meets Wavefunction Realism'', \textit{The British Journal for the Philosophy of Science}. \url{https://www.journals.uchicago.edu/doi/10.1086/720523}.
    \item Calosi, C. and D\"{u}rr, P. 2021. ``The general-relativistic case for super-substantivalism", \textit{Synthese} 199: 13789-13822.
    \item Cameron, R. P. 2014. ``Parts Generate The Whole, But They Are Not Identical To It'', in {\em Composition as Identity}, edited by Donald Baxter and Aaron Cotnoir, pp. 90--107, Oxford: Oxford University Press.
    \item Casati, R. and Varzi, A. 1999. {\em Parts and Places}. Cambridge, MA: MIT Press.
    \item Correia, F. 2022. A General Theory of Location Based on the Notion of Entire Location. \textit{Journal of Philosophical Logic}. At: \url{https://link.springer.com/article/10.1007/s10992-021-09641-5.}
    \item Cotnoir, A. J. and Varzi, A. 2021. {\em Mereology}. Oxford: Oxford University Press.
    \item Crowther, K. 2018. “Inter-theory relations in quantum gravity: Correspondence, reduction, and emergence”, {\em Studies in History and Philosophy of Modern Physics}, 63: 74-85.
    \item Eagle, A. 2016. “Persistence, Vagueness, and Location”, {\em Journal of Philosophy}, 113: 507–532.
    \item Eagle, A., 2019, “Weak Location”, {\em Dialectica}, 73(1-2): 
    \item Fine, K. 2010. ``Towards A Theory of Part'' {\em Journal of Philosophy} 107(11): 559--589. 
    \item Gilmore, C. 2010. ``Sider, the inheritance of intrinsicality, and theories of composition'' {\em Philosophical Studies} 151: 177-197.
    \item Gilmore, C. 2014. Building Enduring Objects Out of Spacetime. In Calosi, C. \& Graziani, P. (eds). \textit{Mereology and the Sciences}. Cham: Springer, pp. 5-34.
    \item Gilmore, C. 2018. “Location and Mereology”, {\em Stanford Encyclopedia of Philosophy}, (Fall 2018 Edition), Edward N. Zalta (ed.), \url{https://plato.stanford.edu/archives/fall2018/entries/location-mereology/}.
    \item Giordani, A. and Calosi, C. 2023. ``Atoms, coms, syllables and organisms'' {\em Philosophical Studies} 180: 1995--2024. 
    \item Humphreys, P. 1997. “How Properties Emerge", {\em Philosophy of Science}, 64(1): 1-17. 
    \item Huggett, N. and W\"uthrich, C. 2013. ``Emergent Spacetime and Empirical (in)coherence'', {\em Studies in History and Philosophy of Science Part B: Studies in History and Philosophy of Modern Physics}, 44: 276--285.
    \item Kleinschmidt, S. 2016. ``Placement Permissivism and Logics of Location”, {\em Journal of Philosophy}, 113: 117–136.
    \item Lam, V., and Wüthrich, C. 2018. “Spacetime is as spacetime does", {\em Studies in History and Philosophy of Science Part B: Studies in History and Philosophy of Modern Physics}, 64: 39-51.
    \item Leonard, M., 2021a. ``Supersubstantivalism and the Argument from Harmony", \textit{Thought} 10 (1):53-57.
     \item Leonard, M. 2021b. ``What is it to be located?'' {\em Philosophical Studies} 178: 2991-3009.
    \item Le Bihan, B. 2018a. ``Space Emergence in Contemporary Physics'', {\em Disputatio}, 49(10): 71-85.
    \item Le Bihan, B. 2018b. ``Priority Monism Beyond Spacetime'', {\em Metaphysica}, 19(1): 95-111. 
    \item Le Bihan, B 2021. ``Spacetime Emergence in Quantum Gravity: Functionalism and the Hard Problem'', {\em Synthese}, 199(2): 371-393.
	\item Loss, R., 2021. ``Somewhere Together: Location, Parsimony and Multilocation.'' \textit{Erkenntnis}. At: \url{https://link.springer.com/article/10.1007/s10670-021-00376-y}.
 \item Loss, R., 2016. ``Parts Ground the Whole and are Identical to it'', {\em Australasian Journal of Philosophy} 94(3): 489--498. 
 \item Lehmkuhl, D. 2018. ``The Metaphysics of Super-Substantivalism'', {\em No\^us} 52:1: 24–4.
\item Lewis, D. 1991. {\em Parts of Classes}. Oxford: Basil Blackwell. 

 \item Mellor, D. H. 2008. ``Micro-compostion'' {\em Philosophy} 83: 65-80.
 \item Ney, A. 2021. {\em The World in the Wavefunction}. Oxford: Oxford University Press.
    \item Parsons, J. 2007. “Theories of Location”, in D. Zimmerman (ed.), {\em Oxford Studies in Metaphysics}, 3: 201–232.
    
    \item Simons, P. 2004. “Location”, {\em Dialectica}, 58: 341–347.
    \item Sider, T. 2007. ``Parthood'', {\em Philosophical Review}, 116: 51--91. 
   \item Smid, J. 2017. ``What does ``nothing over and above its parts'' actually mean?'' {\em Philosophy Compass} 12(1): e12391. 
    \item Varzi, A. 2007. “Spatial Reasoning and Ontology: Parts, Wholes, and Locations”, in {\em Handbook of Spatial Logics}, edited by M. Aiello, I. Pratt-Hartmann, and J. van Benthem, pp. 945–1038,  Berlin: Springer-Verlag. .
    \item Varzi, A. 1996. ``Parts, wholes, and part-whole relations: The prospects of mereotopology, {\em Data \& Knowledge Engineering} 20: 259--286.
    \item Wilson, J. 2021. {\em Metaphysical Emergence}. Oxford: Oxford University Press.
\end{description}

\end{document}